# Multi-synchronous collaboration between desktop and mobile users: A case study of report writing for emergency management

**Adrian Shatte, Jason Holdsworth, Ickjai Lee**
College of Business, Law and Governance (Information Technology)
Division of Tropical Environments & Societies
James Cook University, Cairns, Australia
Email: Adrian.Shatte@my.jcu.edu.au, Jason.Holdsworth@jcu.edu.au, Ickjai.Lee@jcu.edu.au

## Abstract

The development of multi-synchronous decision support systems to facilitate collaboration between diverse users is an emerging field in emergency management. Traditionally, information management for emergency response has been a centralised effort. However, modern devices such as smartphones provide new methods for gaining real-time information about a disaster from users in the field. In this paper, we present a framework for multi-synchronous collaborative report writing in the scope of emergency management. This framework supports desktop-based users as information providers and consumers, alongside mobile users as information providers to facilitate multi-synchronous collaboration. We consider the benefits of our framework for writing collaborative Situation Reports and discuss future directions for research.

## 1  Introduction

Effective knowledge management is fundamental in fast changing environments such as emergency situations (Yates and Paquette 2011). Decision-making in emergency situations requires timely and accurate information from a range of stakeholders including law enforcement, emergency response personnel, and other experts (Ludwig et al. 2013). Traditionally, information management for disaster response has been a centralized effort, however modern technologies such as social media have provided new methods for gaining accurate information in real-time (Yates and Paquette 2011).

Multi-synchronous collaboration is a process in which some users work in real-time (e.g. desktop-based users) while other users work in isolation and commit updates when necessary (e.g. mobile users) (Rahhal et al. 2009). The nature of disaster management in terms of time pressure, uncertainty, complexity, and the need for collaboration between multiple organisations means that emerging decision support systems are required to be real-time, multi-synchronous, and capable of supporting distributed users on different devices. Collaborative report writing is an important task conducted by emergency service agencies during a disaster to collate and distribute the facts of a disaster situation (Schulz et al. 2012). Multi-synchronous collaboration is an emerging area of research that can facilitate report writing for emergency management.

One method for gaining accurate accounts of an emergency situation is to have trusted experts at the disaster site providing updates via mobile devices (Ludwig et al. 2013). The popularity of mobile devices is attributed to the convenience of accessing applications anywhere at any time, independent of the user's location or movements (Fernando et al. 2013). In the realm of emergency management, mobile devices have been used to facilitate communication between offsite personnel (e.g. staff in a control centre) and onsite personnel (e.g. first responders on the scene) (Ludwig et al. 2013). Despite the benefits of mobile devices, there are several challenges inherent with mobile computing including limited memory, finite battery, and unreliable network connectivity (Shiraz et al. 2013). Further, challenges inherent in disaster management including time pressure, uncertainty, complexity, and multiple stakeholders highlight the need for flexible, real-time collaboration (Ludwig et al. 2013).

With mobile devices becoming increasingly ubiquitous, it has been suggested that smartphones can facilitate information sharing and decision making in disaster environments (Monares et al. 2011). However, little research has investigated the use of mobile devices to facilitate multi-synchronous report writing to support decision-making in emergency situations. In this paper we present our work in investigating methods to integrate mobile users with real-time collaboration for emergency document writing. This work culminates in the development of a web-based system to facilitate collaboration between agencies when writing reports for emergency management. The web-based system interfaces with mobile users by providing temporary locks on content and a push-based synchronisation mechanism to commit changes. In this way, mobile users play the role of information providers while desktop-based users are information providers and consumers. Further, standard web technologies provide a consistent and convenient experience to users across all types of devices.

To evaluate the applicability of our system we consider the case of Situation Reports (sitreps). We comment on the usefulness and applicability of our framework for completing sitreps, focusing on benefits and potential limitations. Further, we compare our system to existing solutions that utilise mobile devices for emergency management. As research and development of our prototype is ongoing, we also consider avenues for future work. The remainder of this paper is structured as follows: Section 2 provides a background on emergency management and considers existing systems; Section 3 describes our web-based and mobile prototypes to support collaborative writing for emergency management; Section 4 describes the case of sitreps and evaluates the benefits and limitations of our method in comparison to existing systems; and Section 5 discusses the conclusions and speculates on future work.

## 2 Preliminaries

### 2.1 Communication in Disaster Management

Effective communication and collaboration between key organisations is critical in disaster situations. However, communication is consistently identified as one of the biggest challenges within emergency management. For example, the commission into the 2009 Victoria Bushfires concluded that confusion over leadership and coordination issues impacted the effectiveness of response (Teague et al. 2010). Similarly, a review of the civil response to the 2011 Christchurch earthquake criticised the segregation between organisations in terms of disaster management software and information sharing (McLean et al. 2012).

Unexpected problems, dynamic changes of situations or environmental and knowledge limitations lead to the need for improvisation in crises and emergencies (Stein 2011). For this reason, it is important to ensure that emergency management systems allow for improvisation during a disaster. Based on the analysis of previous disasters, six characteristics were identified that make improvisation important (Ludwig et al. 2013; Mendonça 2007):

- *Rarity of disasters:* it is rare for major disasters to occur which limits opportunities for training and learning in real situations;
- *Time pressure:* the steps of planning and executing solutions are forced to converge due to time constraints in an emergency, meaning that some decisions must be made with limited information available;
- *Uncertainty:* it is difficult to predict how a crisis will develop over time so predictions and improvisation are necessary based on intelligence;
- *High and broad consequences:* often the effects of a large disaster are high and broad, which means that interdependencies among a wide range of physical and social systems must be managed to ensure adequate response;
- *Complexity:* based on the *high and broad consequences* of a disaster, complexity arises;
- *Multiple decision makers:* disasters require the negotiation and collaboration of multiple organisations.

Information and Communication Technologies (ICT) can be employed to assist communication, emergency management processes and improvisation in all phases of emergency management. The main benefits of ICT for emergency management are: 1) diverse communication channels can be utilised for effective warnings and notifications; 2) information from diverse sources can be easily integrated into a single system; 3) disaster relief efforts can be centrally coordinated; 4) systems can encourage both public and institutional input; and 5) damage caused and other effects of a disaster can be analysed to inform recovery and future response (Li et al. 2014).

## 2.2 Existing Collaborative Systems for Emergency Management

Several approaches currently exist for supporting communication and collaboration in the scope of emergency management. Table 1 lists several web-based applications that are specifically tailored for collaborative document writing for emergency management. These applications integrate data from diverse agencies social media to generate documents such as incident reports and sitreps.

| Name of system | Main Features | Reference |
| --- | --- | --- |
| Emergency Response Intelligence Capability (ERIC) | Integrates data from state and federal agencies into a single web-based map interface, historical snapshots, and automated sitreps. | (CSIRO, 2015a) |
| Emergency Situation Awareness (ESA) | Detects unusual behaviour on Twitter to quickly alert the user when a disaster event is being broadcast. | (Cameron et al. 2012; CSIRO 2015b) |
| uEmergency | Interactive tabletop system that displays maps and supports annotations. History slider allows users to view different situations over time. Support single-user, multi-user, and mixed collaboration methods. | (Qin et al. 2012) |

*Table 1. Web-based collaborative systems for emergency management*

The emergence of mobile devices has resulted in distributed solutions for emergency management. Typically, these solutions involve two types of workers: offsite workers (e.g. personnel working in a control centre) and onsite workers (e.g. users at the site of the disaster). The offsite workers are typically the decision makers, whereas the onsite mobile users are information providers (Nilsson and Stølen 2010). Table 2 lists existing solutions for emergency management that include mobile frameworks.

| Name of system | Main Features | Reference |
| --- | --- | --- |
| MoRep | Android application for viewing reports, requesting reports from subordinates, and creating reports with multimedia during an emergency situation. Information is displayed as annotations on a map. | (Ludwig et al. 2013) |
| TwiddleNet | Uses smartphones as personal servers to enable instant content capture and dissemination for first-responders in emergency situations. Automatic tagging of content and distribution to several networking channels. | (Singh and Ableiter, 2009) |
| MobileMap | Complements radio communication to allow for ad hoc communication (when network availability is unreliable), decisions support and collaboration among workers in the field using mobile devices. | (Monares et al. 2011) |

*Table 2. Mobile-based systems for emergency management*

While the systems listed in Tables 1 and 2 provide many diverse features, none of these systems support multi-synchronous collaboration between desktop and mobile users. It is important to support collaboration between centralised and distributed users during an emergency situation to ensure that the right information reaches the right people at the right time.

## 2.3 Challenges in Supporting Multi-synchronous Collaboration

Supporting real-time collaboration between control centre staff and mobile users has several challenges. While control centre staff may have access to a reliable network, power, and computing resources, mobile users may not have these benefits (Shiraz et al. 2013). Further, one of the biggest challenges in internet-based collaboration is maintaining a consistent view for all collaborators (Koren et al. 2013). Synchronisation techniques, such as Differential Synchronisation (diffsync), are robust to high latency or unreliable network environments and are naturally convergent (Fraser 2009). In fields such as emergency management where accurate and timely information is imperative, diffsync can provide robust synchronisation, but additional techniques should be investigated to facilitate collaborative multi-synchronous report writing with input from first responders in the disaster area.

# 3  Framework for multi-synchronous collaborative report writing

Based on the need for multi-synchronous collaborative report writing in emergency management, we developed a prototype system to support distributed collaboration between browser-based and mobile-based users. This development is described in this section of the paper. First we outline the research methods that guide both the implementation and evaluation stages of our project. Next we describe the web technologies that were used in the development of our prototype for collaborative report writing in emergency situations. Finally, we provide an overview of the prototype with specific focus on its major features and how these features intend to support multi-synchronous collaboration between web and mobile users.

## 3.1  Research Methods

Our ongoing research into collaborative report writing for emergency management is guided by three main research methods. The first method is the *Design Science* approach (Von et al. 2004). We have developed several features for web-based collaboration including flexible locking and user attribution, and the results of these studies have guided our current work in developing a multi-synchronous framework. Additionally, our work utilises several tools from disparate sources (described in section 3.2) that we have adapted for use in our prototype. As such, we follow the principles of *Mashup Development (Yu et al. 2008)* which is a methodology for building new web applications based on existing data and frameworks. Thirdly, we follow a *Toolkit* approach - the proposed framework extends our previous work on the diffsync technique to develop collaborative editing features that can be plugged into existing applications.

## 3.2  Framework

Figure 1 displays the framework of our system for multi-synchronous collaboration between desktop-based users and mobile users for collaborative report writing. Multiple desktop-based clients can be connected to the system for real-time editing. These clients utilise the diffsync synchronisation mechanism to maintain consistency of shared text as changes are being made in real-time. Meanwhile, mobile clients are provided with small tasks via a locking mechanism. The mobile clients do not collaborate in real-time, instead pushing changes back to the main document as needed or when a network connection is available. Using this framework, we developed a prototype system.

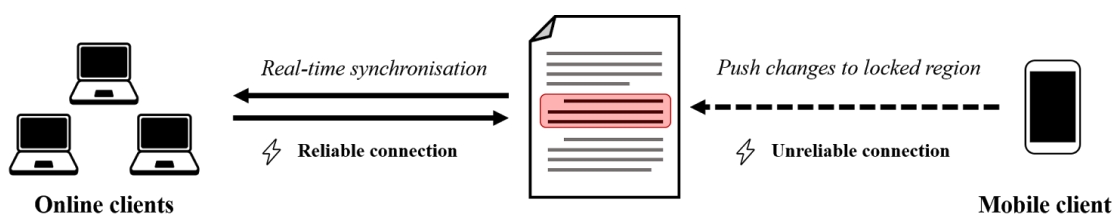

*Figure 1: Framework of our system. Online users converge with real-time synchronisation whereas mobile users push updates to isolated sections of the document.*

## 3.3  Development tools used in our prototype

The following software was used to develop our prototype:

### 3.3.1  Diffsync

For synchronisation of the shared text between collaborators our prototype uses the diffsync technique (Fraser 2009). Diffsync is a client-server technique for plain text document synchronisation with open source implementations written for common programming languages, including JavaScript (Lautamäki et al. 2012). We selected diffsync for our prototype due to its ease of implementation, high responsive, scalability, and natural convergence (Fraser 2009). Further, diffsync is designed so that it can be appended to existing applications with relative ease, thus introducing real-time collaboration to previously static systems (for example, MobWrite (Fraser 2010)).

### 3.3.2 Node.js

Node.js is a highly scalable, server-side JavaScript environment based on Google's V8 runtime engine (Tilkov and Vinoski 2010). It is commonly used to develop real-time applications (such as web servers and networking tools) due to its single-threaded, asynchronous nature. Node.js is also non-blocking and support concurrency, which is important for multi-synchronous web applications with high request demands. All of these features make useful for developing real-time, collaborative text-editing systems. Based on the original specifications presented by Fraser (Fraser 2009), we implemented Guaranteed Delivery diffsync in Node.js.

### 3.3.3 Ace Editor

Ace editor is a web-based text editor that is compatible with JavaScript applications. Ace editor provides many features that can improve web-based document writing include highlighting, indentation, line numbers, and annotations. Additionally, Ace is capable of supporting both small and large documents, which makes it useful for document scalability. We selected Ace editor to exploit its highlighting functionality for mobile locking and also for its compatibility with the other tools.

### 3.3.4 Google Maps API

The Google Maps JavaScript API provides public map features for web-based applications. Key features include road map and satellite imagery, integrated data from diverse sources (e.g. business information, landmarks, etc.), customisation, and advanced visualisation features (e.g. heatmaps, markers, and symbols). We selected Google Maps for our prototype to display the position of offsite users on mobile devices. As the Google Maps API is written in JavaScript it is compatible with our other selected technologies.

## 3.4 Web-based and mobile applications

Our document writing prototype consists of two separate applications: 1) a web browser-based component, and 2) a mobile component. In this section of the paper we describe the intended workflow of users within our application and also describe the interfaces and functionality of the browser-based and mobile components.

The intended workflow of our application is as follows:

- Emergency staff are working on a report at a control centre. The web-based, real-time collaborative nature of the system allows for several individuals and agencies to collaborate in a single shared workspace.

- Meanwhile, remote emergency staff are moving throughout the disaster area equipped with mobile devices (e.g. smartphones or tablets). Control centre staff can view the location of these users on a map within the report writing system. When information about the situation is required, tasks can be assigned to mobile users in which information is requested about their immediate surroundings such as condition of roads or flooding.

- The mobile users are notified when a new task is received. Notes can be added to the task and synchronised with the server when ready.

- Meanwhile on the server, online users can continue to edit the document in real-time. They will see that certain sections are locked for remote users, and they won't be able to make any changes to those areas to avoid conflicts. It does not matter what changes are made within other sections of the document as the underlying synchronisation method ensures that locked sections adjust position automatically when necessary.

- When a remote worker has network access and is ready to synchronise information, synchronisation can be triggered within the smartphone application. This synchronisation is different to normal diffsync as only the locked content is pushed to the server and replaces the current content within the relevant lock.

- In this way, both desktop-based users and mobile users with limited resources can work together in a multi-synchronous environment while maintaining consistency.

The web-based component of our prototype is displayed in Figure 2. The interface has several key features. First, there is a large section used for collaborative report writing. Templates can be used so that a new sitrep can be started efficiently. All collaborators using the web-based interface will see the same text, due to synchronisation happening in real-time. Second, there are three sub features used to provide tasks to mobile users in the disaster area:

1) A user can select an area of text and use the "Lock Mobile" button at the top of the interface to push a new task to a mobile user. The user is prompted to enter a brief description of the task which is sent to the mobile user.

2) Current mobile tasks are displayed in a list to the right of the text-editing screen. Each task is identified with both a colour and a username. The system provides each mobile user with a unique colour so that each task and the related section of the document can be highlighted as necessary. The task also shows an indication of when the mobile user last updated their text (e.g. *3 mins ago*). A task can be cancelled from this view as well.

3) A map displays the last known location of a mobile user in the field. This information is updated whenever a mobile device synchronises with the web-based system, which occurs automatically every five minutes (pending network availability) or whenever text is manually synchronised by the mobile user. This provides web-based users with awareness of the mobile users' locations, aiding in decision-making when assigning tasks to mobile users.

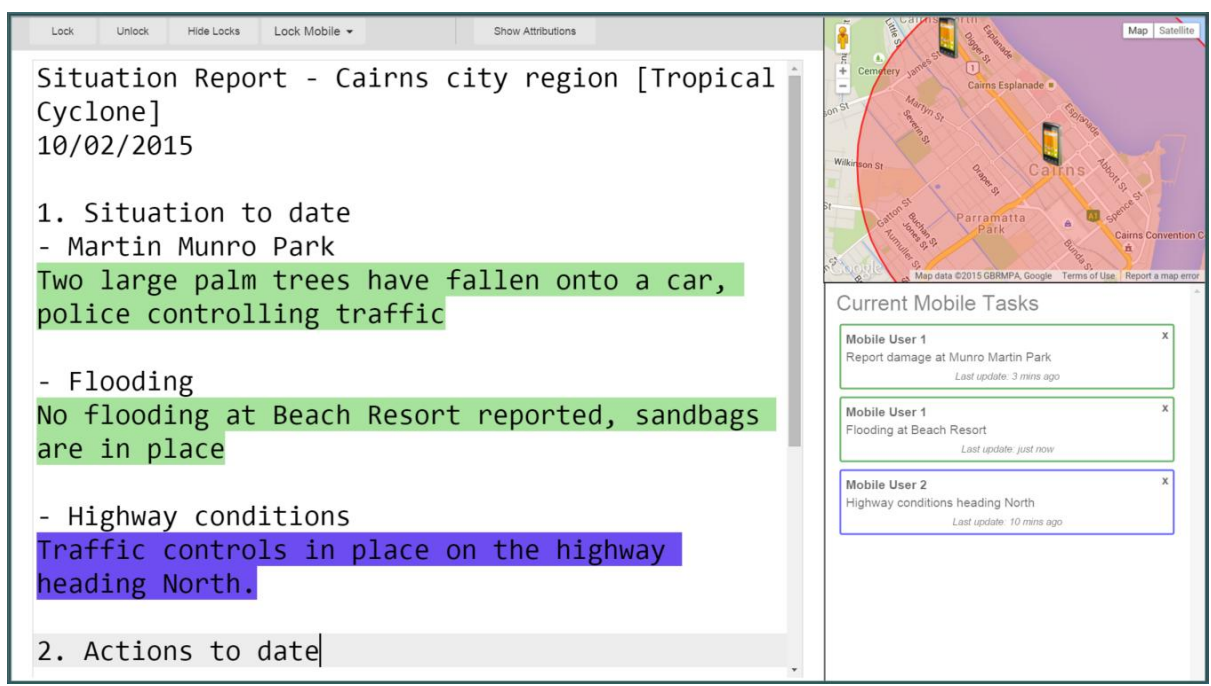

*Figure 2: The web-based interface of our system showing several tasks locked to mobile users.*

## 3.5 Mobile Features

The second component of our prototype system is a mobile application for information providing tasks. This application allows mobile users to receive requests from centralised users to provide information about a context-specific incident or situation within the disaster area. Screenshots of the mobile interface are displayed in Figure 3a, Figure 3b, and Figure 3c.

Notifications are received by the mobile device when a new task is locked to that particular user from the web interface (see Figure 3a). A user can click directly on the notification to enter immediately into the note taking screen (Figure 3c). Notifications are also received when a web user revokes the mobile lock (e.g. if no further information is required from the mobile user). This ensures that the mobile user is aware that no further information is required and will not spend any more time on this task.

All tasks that are currently assigned to the mobile user are displayed in the main screen of the application (Figure 3b). For each task, a brief description of the task and the time it was assigned is displayed. A user can click on a specific task to enter the note taking screen (Figure 3c). There is also the option to refresh the list of current tasks (triggered in the submenu).

The note taking screen (Figure 3c) displays the brief description of the information required at the top of the screen. The synchronisation button is used to push information back to the main document, and an indication of when the last synchronisation occurred is also displayed. The main text area below these controls is where the mobile user adds information about the disaster situation as necessary. The mobile user can also dismiss a task from the submenu of this screen to notify the web users that no more information will be provided about this task.

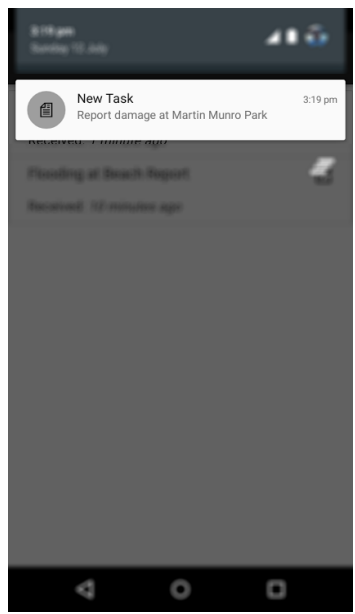 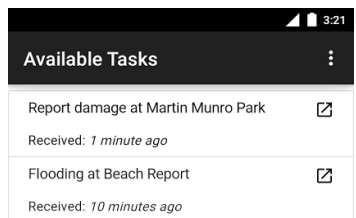 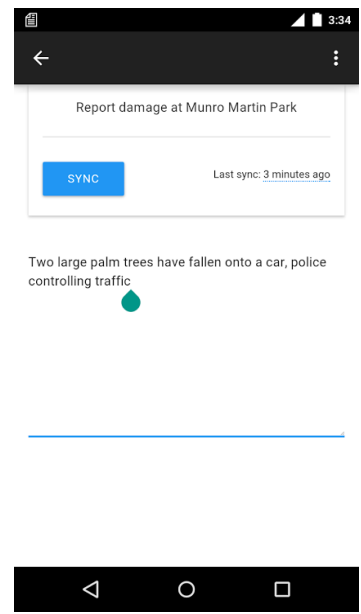

*Figure 3a: Notifications are received when a new task is available.*

*Figure 3b: The user can view a list of all active tasks they are assigned.*

*Figure 3c: Notes can be added about the task and synchronised.*

## 4 Scenario: Situation Reports

To consider the benefits and applicability of our collaborative prototype in report writing for emergency management, we consider the Situation Report (sitrep) document. In this section of the paper we describe the details of sitreps, potential weaknesses in the process of writing sitreps, and discuss benefits of our prototype at overcoming these challenges with specific focus on benefits of mobile locking for multi-synchronous collaboration.

### 4.1 Definition of Sitrep

The sitrep is a document used by emergency management personnel during a disaster situation to facilitate decision-making. The sitrep records an ongoing description of all events and incidents occurring during an emergency situation (e.g. an extreme weather event). The content of a sitrep must be both accurate and timely, as these details are passed on to adjacent agencies or higher headquarters and Emergency Operations Centres.

The structure of a sitrep can be flexible and there are many suggested formats. For example, the Australian government suggests that sitreps contain details such as a report version number, date and time, type of incident, location of incident, contact details of the relevant incident management centre, casualties (including dead, injured, evacuated, and homeless), general details about the situation and damage, actions in progress, assistance required, future intentions, and the overall prognosis of the situation (Emergency Management Australia 1998). Other guidelines state that a sitrep should contain factually verified information only, be brief, and should be specific to a single functional area (e.g. roads). This means that during an emergency there may be multiple sitreps in use to cover different functional areas. Sitreps can also include graphics such as maps.

While sitreps can be useful for providing brief but informative facts about an emergency incident, there can be challenges in using them effectively. Several reports have noted that lack of effective coordination, poor communication and confusion over situation awareness result in ineffective sitreps (McLean et al. 2012; Teague et al. 2010). In light of the challenges and shortcomings of sitreps and ICT in previous disaster situations (McLean et al. 2012; Teague et al. 2010), we considered how our use of framework for multi-synchronous collaboration could benefit collaborative sitrep writing.

### 4.2 Benefits and Limitations of Multi-synchronous Collaboration for Sitreps

One of the key benefits of using a multi-synchronous approach to collaborative sitrep writing is that accurate information can be provided by trusted workers in the field. When accuracy is important, information provided by the public needs to be verified before it can be used to inform disaster response. However, recent research indicates that large groups may perform as well or better than individual experts in producing useful information (Vivacqua and Borges 2012). Our framework could easily be extended to engage public users through mobile devices for situation reporting.

Another benefit is that mobile users can move throughout a disaster environment, whereas workers in a control centre are at a fixed location. This means that as the conditions change in the disaster, mobile users can move around. According to Guerro et. al. (2006), handheld devices are considered more appropriate than laptops or tablets in unsafe environments (e.g. a disaster site). This is because they are easy to deploy and carry, require lower user attention, and have a short startup time. These features combine to allow for quick reactions from a user, which could be necessary in volatile environments.

A mobile user wandering around an environment will likely engage in short and simple interactions with the system, thus only basic communication support is required (in terms of network availability and bandwidth). In contrast, a user in a fixed position requires larger amounts of bandwidth and greater network availability, as interaction with the system is typically longer and more complex (Guerrero et al. 2006). This is the intended use case for our framework, as control centre staff are considered as information providers, consumers, and decision-makers, whereas mobile users are only required to contribute small sections of information about a single incident.

There are also some limitations we identified. For instance, as mobile users are given single, isolated tasks, there is a lack of situation awareness and their responses cannot be tailored to the context of the wider document. This is mitigated through the use of a task description, but additional mechanisms such as radio or chat could be used to clarify details (Ludwig et al. 2013). Further, some aspects of the communication are one-sided. Staff within a control centre can push notifications to mobile users for information, but there is no method within our system for mobile users to initiate communication or submit content without a lock first being granted.

It should also be noted that our current system has a heavy dependency on a stable network connection. While diffsync is robust to network faults such as packet loss or data corruption (Fraser, 2009), complete loss of Internet connection would render the system not operational. For this reason, it is important to consider alternative network solutions that could support our system in an emergency situation. Nilsson and Stølen (2010) distinguish four main network solutions for emergency response: 1) wireless ad hoc networks; 2) cellular networks; 3) special emergency networks; and 4) router-based networks that are deployed for the emergency operation. They conclude that wireless ad hoc networks are well-suited for emergency response as they are quick to deploy, provide local communication within the disaster area, and can be used alongside an Internet connection (if it is available) (Nilsson and Stølen 2010). In the event that no alternative networks are available, sitreps could still be completed collaboratively with field workers relaying information to remote offices using radio or satellite phones.

### 4.3 Comparison to Existing Systems

In terms of report creation, ERIC integrates data from several agency feeds to automatically generate sitreps (CSIRO 2015a). Further, MoRep, and TwiddleNet collect additional information including multimedia from social networks and mobile users (Ludwig et al. 2013; Singh and Ableiter 2009). Our framework utilises information provided by trusted users (e.g. emergency workers) through either the web or mobile interfaces. This information could be augmented with automated feeds and social media to provide situation awareness to workers in the shared document and improve decision-making.

Other systems including ESA utilise social media trends and data mining techniques to detect disaster events as they occur (CSIRO 2015b). Our framework could benefit from these features, in the sense that control centre staff (e.g. web users) could monitor social media trends to inform the types of tasks that are requested from mobile users. For example, if there are numerous social posts about a fire within the disaster area, information about possible files can be requested from the mobile users. In this sense, situations would be crowdsourced from social media and verified by trusted emergency service workers.

With uEmergency and MobileMap, face-to-face or radio communication is enhanced by the use of an interactive, collaborative system (Monares et al. 2011; Qin et al. 2012). Communication between mobile users and web-based users could be facilitated using a secondary source such as radio or phone. This may be necessary if a mobile user requires additional context to complete an assigned task. However, public channels can become crowded with noise in a disaster situation (Ludwig et al. 2013), so future development might consider integrating dedicated communication features into the application itself.

Finally, there could be issues with reliable network connectivity in disaster situations. MobileMap utilises an ad hoc network between workers in the field to share information when there is no network available (Monares et al. 2011). In contrast, our application uses a manual push-based synchronisation mechanism for mobile users, which means that content can be generated while offline and pushed to the server when a connection is restored.

## 5 Conclusion

In this paper we considered the benefits of a framework to support multi-synchronous collaboration between mobile and web users for emergency document writing. Our framework allows web-based users to act as information providers and consumers while mobile users act as information providers. Specifically, we focused on the benefits of this framework for the topic of collaborative sitrep writing. We cite some of the past challenges with ICT systems and sitreps for emergency management and consider how our framework could overcome these challenges and support better information gathering in disaster environments.

After comparing our framework with existing systems for emergency management, we have identified several areas for future research. Firstly, our mobile system could be extended to support crowdsourcing of information from the general public. Second, integration with social media and automated data mining techniques could improve the allocation of tasks to mobile users and provide additional insights to support decision-making. Third, techniques to support network connectivity between mobile and web users should be further investigated to ensure that collaboration can continue when Internet connectivity is disrupted. Interviews with real emergency field workers on the efficacy of our prototype will provide further insights into the usefulness and applicability of our prototype as well as guide further development.

To conclude, our research in this area has indicated that collaboration between web-based and mobile users may benefit decision-making in document writing for emergency management. We intend to build upon this research in the future to investigate the usefulness of a system in other areas of emergency management. We also plan to incorporate additional features and improve on the technical aspects of the system.

# Copyright